\documentclass[a4paper,twocolumn,prb]{revtex4}
\usepackage{amsmath}
\usepackage{graphicx}
\usepackage{color}
\usepackage{subfigure}
\usepackage{hyperref}
\begin{document}
\title{Bak-Tang-Wiesenfeld Model in the Finite Range Random Link Lattice}
\author {M. N. Najafi}
\email{morteza.nattagh@gmail.com}
\affiliation{Department of Physics, University of Mohaghegh Ardabili, P.O. Box 179, Ardabil, Iran}
\begin{abstract}
We consider the BTW model in random link lattices with finite range interaction (RLFRI). The degree distribution for nodes is considered to be uniform in the interval $(0,n_0)$. We numerically calculate the exponents of the distribution functions in terms of $(n_0,R)$ in which $R$ is the range of interactions. Dijkstra radius is utilized to calculate the fractal dimension of the avalanches. Our analysis shows that there is, at least one length scale ($r_0(n_0,R)$) in which the fractal dimension is changed. We find that for the scales smaller than $r_0(n_0,R)$, which is typically one decade, the fractal dimension is nearly independent of $n_0$ and $R$ and is equal to $1.4$, i.e. close to that of the BTW in the regular lattice ($1.25$). Using this fact and other analysis, we conclude that the BTW-type behaviors are dominant for small values of $n_0$ and $R$, whereas for large values of these parameters a new regime is seen in which the exponent of distribution function of avalanche masses is nearly $1.4$. We also numerically calculate the explicit form of the \textit{number of unstable nodes} (NUN) as a time dependent process and show that for regular lattice it is (up to a normalization) proportional to a one dimensional Weiner process and for RLFRI it acquires a drift term. Using this dynamical variable it is numerically shown that we can not continuously approach the regular lattice limit by decreasing $R$.
\end{abstract}
\maketitle
\section{Introduction}
 The concept of self-organized criticality (SOC) was proposed by Bak, Tang and Wiesenfeld (BTW) \cite{BTW} as a possible general framework for explanation of the occurrence of robust power laws in nature which does not require fine tuning of any parameter to set criticality. Unlike the ordinary critical systems, these systems may be open and dissipative and energy input is necessary to offset the dissipation. Sandpile models was the first example of these systems. Despite its simple description, BTW has various interesting features and numerous work, analytical and computational, has been done on this model. Among them one can mention different height and cluster probabilities \cite{MajDhar1}, the connection of the model to spanning trees \cite{MajDhar2}, ghost models \cite{MahRuel}, q-state Potts model \cite{SalDup} avalanche distribution \cite{Avalanche}. For a good review see \cite{Dhar2}. Among the numerous realizations of this model, application of BTW to random link lattices has attracted a lot of attention in recent years. The main motivation for such study was the important observation of Beggs et. al. in which it was shown that the propagation of spontaneous activity in cortical networks is self-organized critical phenomena described by equations that govern avalanches in BTW \cite{Beggs}. The focus of researches in this area had been on the structural and functional properties of random lattice models. One of the most challenges in these systems is finding the circumstances under which the system shows the critical behaviors \cite{Friedman}. In order to monitor critical behaviors, different time series are usually analyzed in which power-law behavior is expected \cite{Herrmann}. Therefore one may be encouraged to investigate time series of topplings in the SOC model on random link lattices.\\
In spite of many theoretical works in this area \cite{kim,kim2,herrmann,holyst} some important issues are missing in the literature such as the effect of finite range of spatial connections (interactions) and the explicit form of distribution function of the topplings $n$ occurred at time $t$, i.e. the statistics of \textit{number of unstable nodes} (NUN) $P(n,t)$. It can help to better understand the dynamics in such systems. We can also approach the regular lattice limit by decreasing the range of interactions and compare the results for two limits, i.e. scale-free random lattice and regular lattice.\\

Here we apply BTW model to random link lattices with finite range interaction (RLFRI). We consider the degree distribution to be $p_d(k)=\frac{1}{n_0}\Theta (n_0-k)$ in which $k$ is the degree of nodes and $n_0$ is degree cut-off (in other words $z_i\leq n_0$ in which $z_i$ is the degree of $i$th node). Although the spatial dimension does not make sense in a completely scale-free networks, it is well defined in RLFRI due to its spatial correlation of links. We first investigate the geometrical properties of the BTW model in RLFRI, i.e. the fractal dimension and the distribution functions of gyration radius and the mass of the avalanches. To this end we define the Dijkstra radius of a clusters (avalanches) \cite{Dijkstra}. We see that there is a length scale ($r_0$) under which the fractal dimension $D_f$ is close to the fractal dimension of avalanches of the BTW model. Observing that $D_f$ is independent of $(n_0,R)$, we can consider $r_0$ as the correlation length at which a smooth change is observed from the BTW-type behaviors to a new regime in which the fractal dimension immediately grows \footnote{Since in our analysis this interval is less than a decade, a reliable amount of fractal dimension can not be reported.}. The properties of the geometrical quantities of the problem are mainly governed by $r_0$. The dependence of this scale to $(n_0,R)$ is studied in this paper. \\

The other important dynamical quantity which we investigate is the NUN as a time dependent process. Noting that approaching to the regular lattice is via reducing the range of interactions between nodes ($R$), one can track the change of dynamics from the free scale lattice ($R\sim \text{system size}$) to the regular one ($R\sim \text{lattice constant}$ and $p_d(k)=4$). We see that the distribution function of NUN at the time $t$, i.e., $P(n,t)$ (in which $n$ is the NUN) for regular (square) lattice and RLFRI has the same symmetry, but their exact formulas are different. We observe that $P(n,t)$ in small $R$ limit does not tend to the regular square lattice meaning that one can not continuously go from the random lattice to the regular one.\\
The paper is organized as follows: in section \ref{ASM} we introduce the problem. In section \ref{gyr} using the Dijkstra algorithm, we determine the radius of the clusters and obtain numerically the fractal dimension of clusters and the distribution function of cluster mass. Section \ref{NUN} is devoted to numerical analysis of NUN.

\section{Abelian Sandpile Model in RLFRI}\label{ASM}
\begin{figure}
\centerline{\includegraphics[scale=.40]{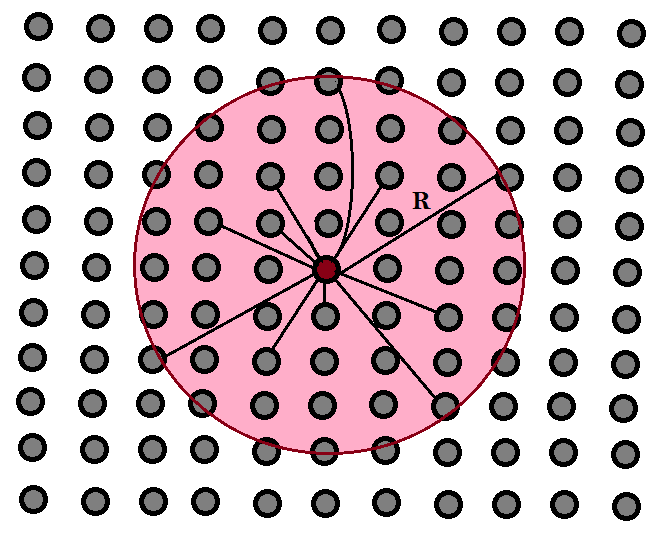}}
\caption{The typical sample showing the model defined in the text.}
\label{sample}
\end{figure}

Let us put the nodes of neural network on a random link $L\times{L}$ square lattice in which each site can be connected to another site inside a disc of radius $R$. Figure \ref{sample} shows schematically the situation. Let the degree of a typical site $i$, $z_i$ (the number of links connected to it) be a random number, conditioned to be in the interval $0\leq z_i\leq n_0$ \footnote{Regarding the fact that for scale-free networks, the degree distribution has the form $p_d(z)\sim z^{-\gamma}$ in which $\gamma$ is the degree exponent, our network model may be interpreted as scale-free network with $\gamma=0$ restricted to the interval $(0,n_0)$.}. We assign a height variable $h_i$ to a typical site $i$ and interpret it as the number of grains in that site. Each $h_i$ is chosen from the set $\lbrace{1, 2,...,z_i}\rbrace$. This yields a configuration of the sand pile which is given by the set $\lbrace{h_i}\rbrace$. Suppose an initial configuration of the pile. We add a grain to a random site $i_0$ i.e. $h_{i_0}\rightarrow{h_{i_0}+1}$, then if the resulting height becomes more than $z_{i_0}$, it becomes unstable and  topples and loses $z_{i_0}$ sands, each of which is transferred to its neighbors. Thus the neighboring sites may become unstable and topple and a chain of topplings may happen in the system. The toppeling in the boundary sites causes some sands leave the system. This process continues until the system reaches a stable configuration. Now one more sand is released to another random site and the process continues. The movement on the space of stable configurations lead the system to fall into a subset of configurations after a finite steps, named as the \textit{recurrent states}. The local toppling rule (occurred in the site $i$) is defined by $h_j\rightarrow h_j+\Delta_{i,j}$ in which:
\begin{equation}
\Delta_{i,j}=\left\lbrace \begin{array}{cc} +1 & \text{if $i$ and $j$ are connected} \\  -z_i & \text{if $i=j$} \\ 0 & \text{other}\end{array}\right.
\end{equation}
For a regular square lattice ($z_i=4$ and $R=1$ for all $i$) it has been shown that the total number of recurrent states is det$\Delta$ where $\Delta$ is the discrete Laplacian. For review see \cite{Dhar3}. It has been shown that the action corresponding to this model is
\begin{equation}
S=\int{d^{2}z}(\partial\theta\bar{\partial}\bar{\theta})
\label{ghost}
\end{equation}
where $\theta$ and $\bar{\theta}$ are complex Grassmann variables. This model corresponds to $c=-2$ conformal field theory.
\section{Gyration radius, Loop and Mass}\label{gyr}
Since we deal with the graphs in RLFRI, the radius of a cluster should be defined in a clever manner. To this end we should first define the boundary of an avalanche and then define its distance from the injection point. Let us define the boundary sites as the toppled sites which have at least one connected untoppled site. To define the distance of two nodes we use the Dijkstra algorithm \cite{Dijkstra}. This algorithm yields the length of the shortest path between two nodes of a graph. According to this algorithm to calculate the distance between two typical nodes $a$ and $b$, one should perform the following steps:\\
\\
(I) Assign to every node a tentative distance value. Set it to zero for the node $a$ and to infinity for all other nodes.\\
(II) Mark $a$ as \textit{current node}, and other as \textit{unvisited nodes}.\\
(III) Calculate the tentative distances for all of unvisited neighbors of the current node. Let us name the current node A and suppose that its distance from $a$ is $d_A$. The new tentative distance of its neighbor B from $a$ (assuming its old tentative distance is $d_B$) will be $\text{Min}\left\lbrace d_A+1,d_B\right\rbrace$. Note that the node B remains in the unvisited set yet.\\
(IV) After considering all of the neighbors of A, mark it as visited and it will never be checked again. If the node $b$ has been marked visited then the algorithm has finished.\\
(V) Among the neighbors of A find the unvisited node with the smallest tentative distance, and set it as the new \textit{current node} and go back to step (III).\\
\\
\begin{figure}
\centerline{\includegraphics[scale=.40]{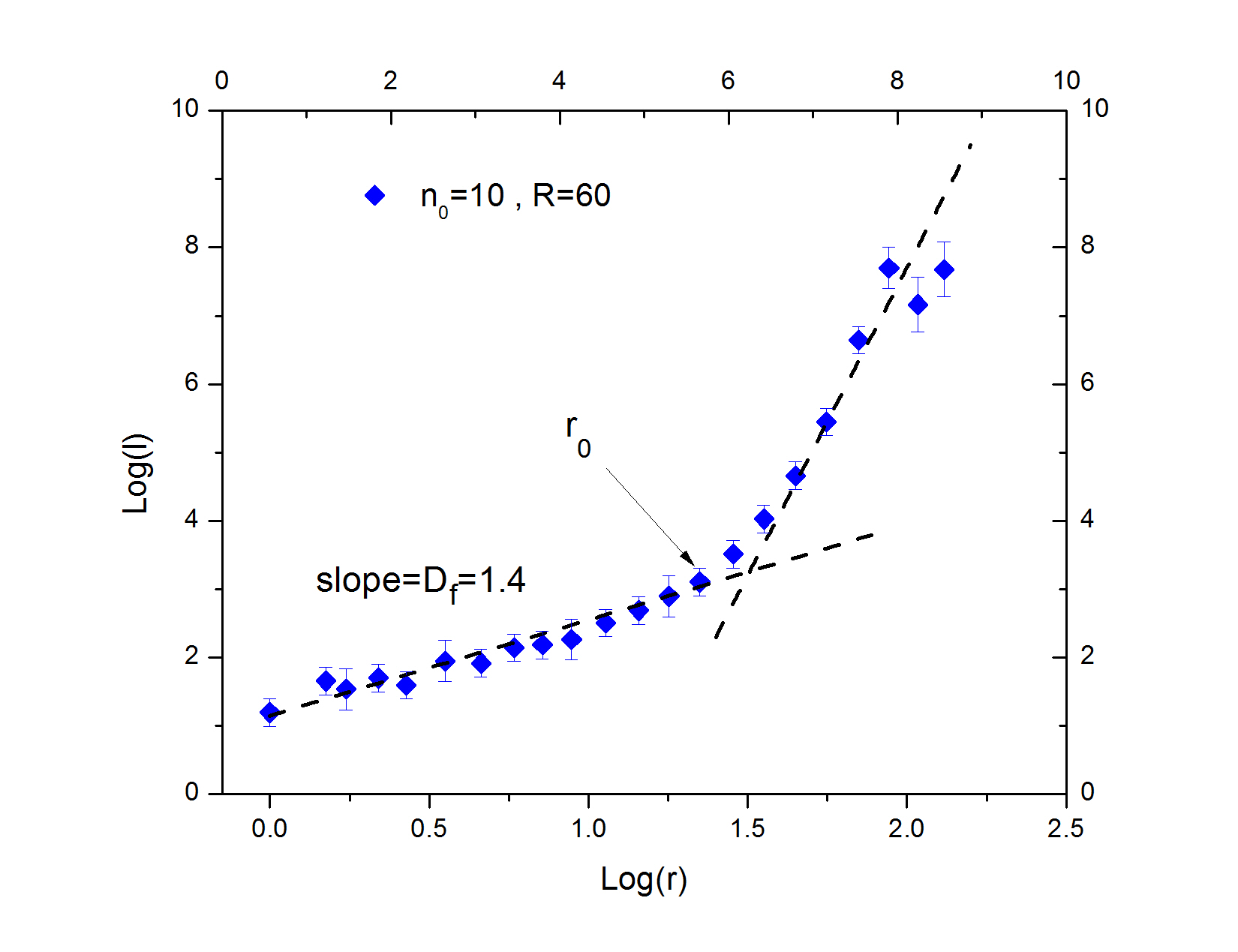}}
\caption{ The fractal dimension of avalanches. It is defined as $\left\langle \text{log}(l)\right\rangle=D_f\left\langle \text{log}(r)\right\rangle$.}
\label{FD}
\end{figure}

\begin{figure}
\centerline{\includegraphics[scale=.35]{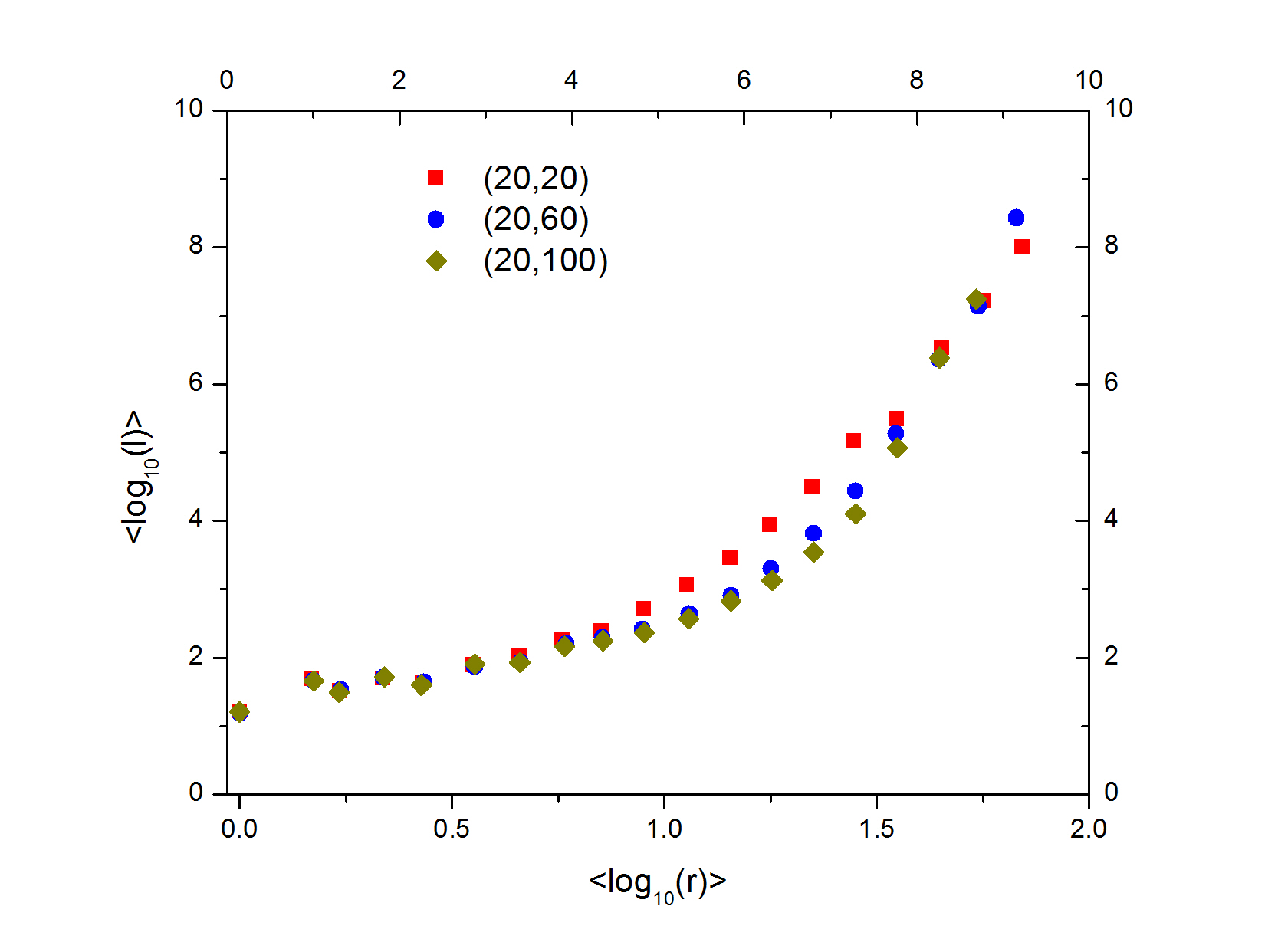}}
\caption{The dependence of $\left\langle \text{log}(l)\right\rangle$ to $\left\langle \text{log}(r)\right\rangle$ for $n_0=20$ and $R=20,60,100$.}
\label{FD2}
\end{figure}

\begin{figure}
\centerline{\includegraphics[scale=.35]{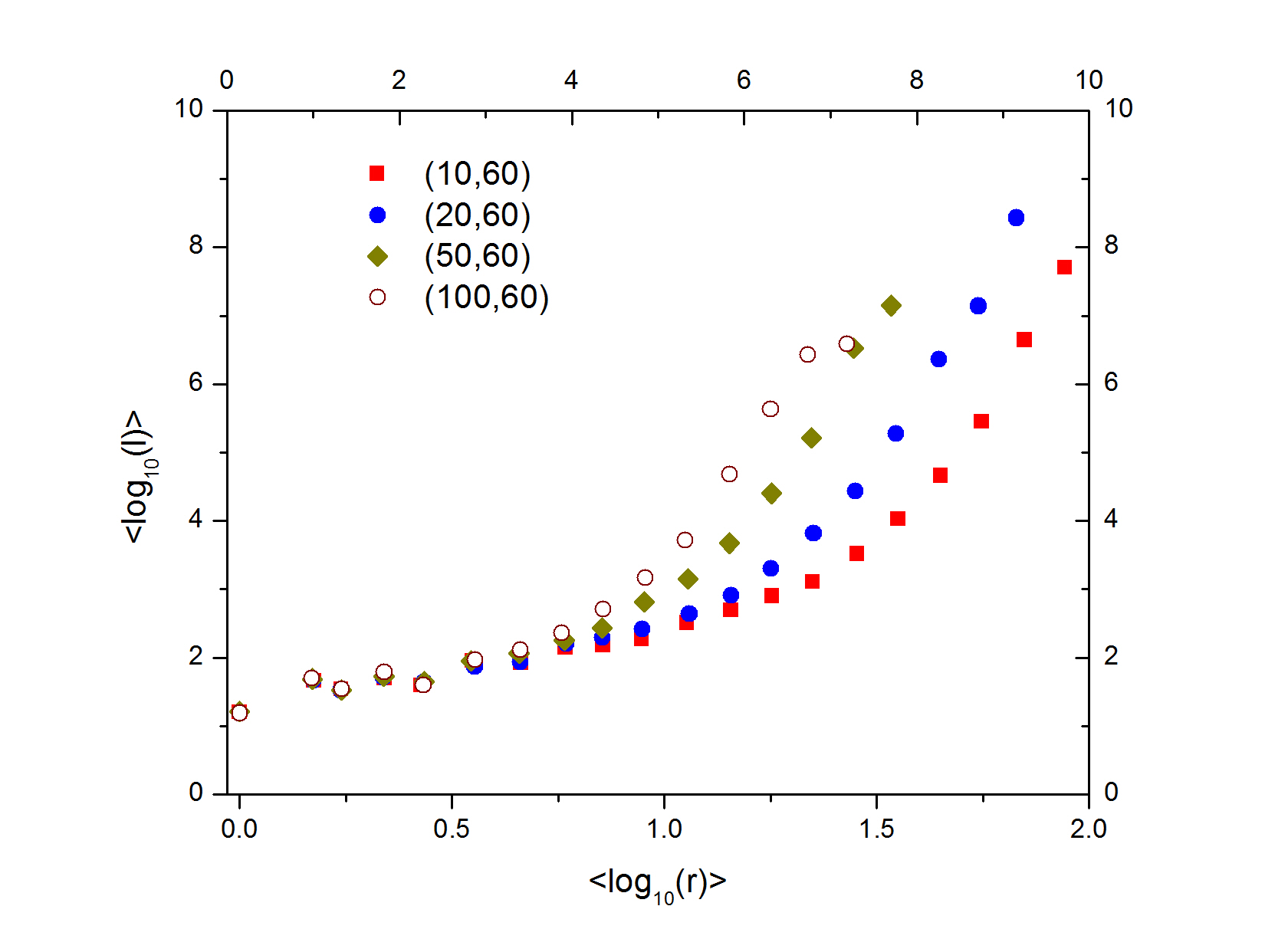}}
\caption{The dependence of $\left\langle \text{log}(l)\right\rangle$ to $\left\langle \text{log}(r)\right\rangle$ for $n_0=10,20,50,100$ and $R=60$.}
\label{FD3}
\end{figure}

\begin{figure}
\centerline{\includegraphics[scale=.35]{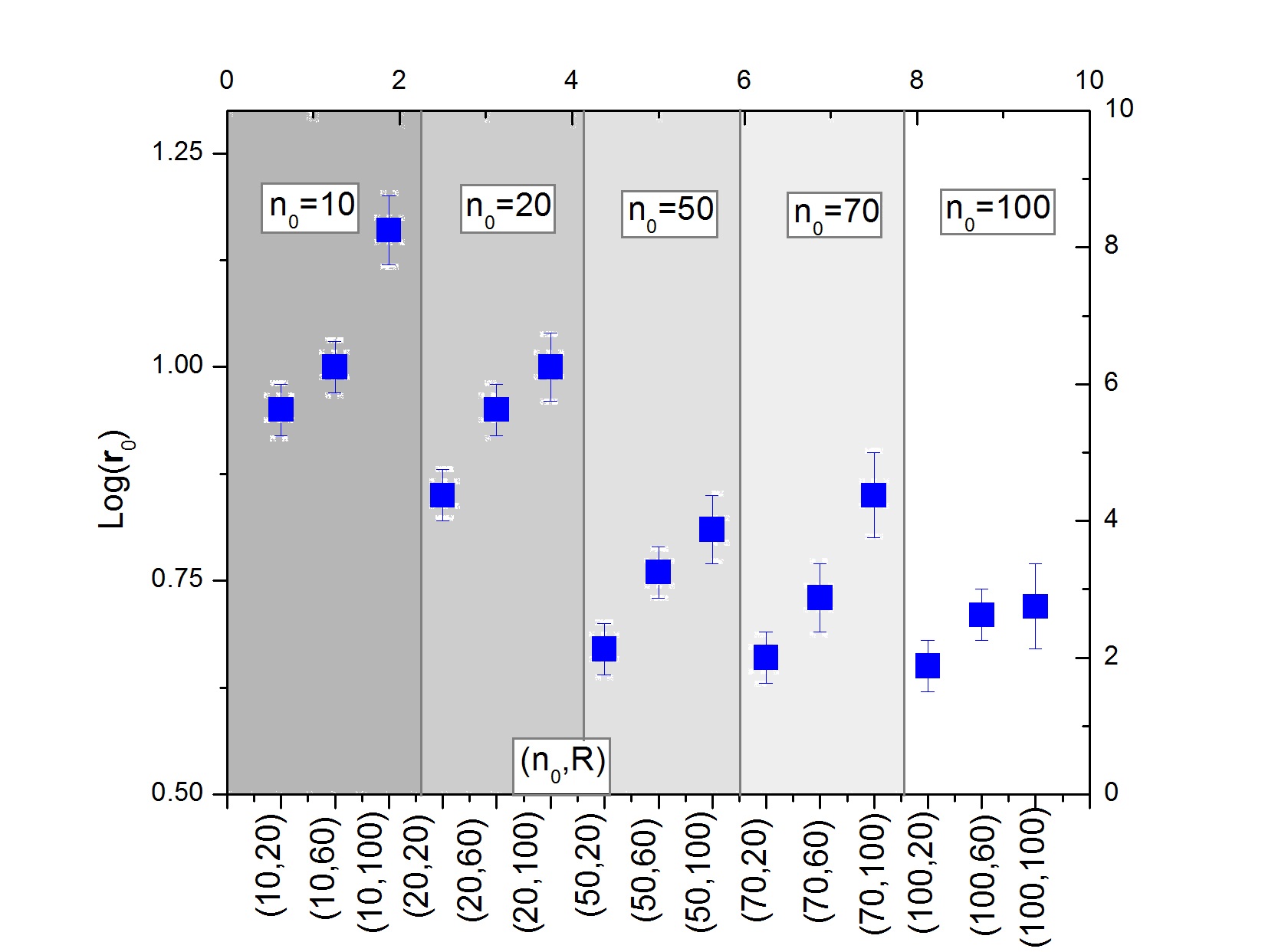}}
\caption{The graph of $r_0$ in terms of $(n_0,R)$.}
\label{FD-scales}
\end{figure}
\begin{figure}
\centerline{\includegraphics[scale=.40]{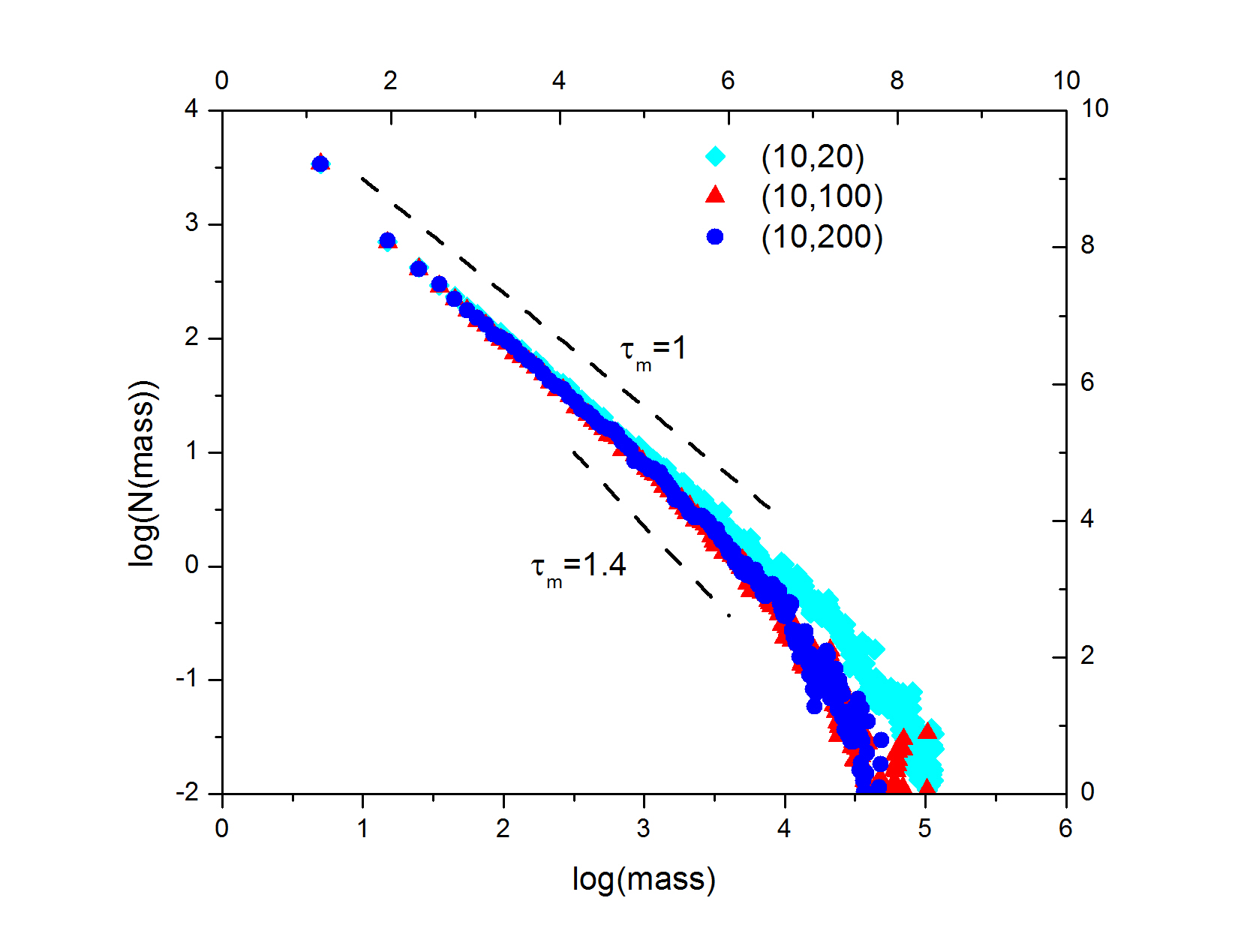}}
\caption{The mass distribution of avalanches.}
\label{mass}
\end{figure}
\begin{table}
\begin{center}
\begin{tabular}{|c|c|}
\hline $(n_0,R)$ & $\log(r_0)(\pm 0.1)$  \\ 
\hline $(10,20)$ & $0.95$ \\ 
\hline $(10,50)$ & $1.0$ \\ 
\hline $(10,100)$ & $1.16$ \\ 
\hline $(20,20)$ & $0.85$ \\ 
\hline $(50,20)$ & $0.67$ \\ 
\hline $(70,20)$ & $0.66$ \\ 
\hline $(100,20)$ & $0.65$ \\ 
\hline
\end{tabular}
\caption{The amount of $r_0$ for various rates of $(n_0,R)$.}
\end{center}
\label{table1}
\end{table}
For this simulation we have considered $512\times 512$ RLFRI and $10^4$ samples are generated for each ($n_0,R$) pair. To begin, we have generated a RLFRI corresponding to some ($n_0,R$) and then a random height has been assigned to each site of the constructed lattice. Approximately after $n\simeq L_x\times Ly$ steps  of adding grains (to random sites), the system reaches a steady state. After each relaxation, we have an avalanche whose boundary has $l$ nodes (boundary length) with the gyration radius (with respect to the injection point) is $r$. In this way one can define the fractal dimension of avalanches by $\left\langle \text{log}(l)\right\rangle=D_f\left\langle \text{log}(r)\right\rangle$. Figure \ref{FD} shows the dependence of $\left\langle \text{log}(l)\right\rangle$ to $\left\langle \text{log}(r)\right\rangle$ for the case $(n_0,R)=(10,60)$. Two distinct behaviors are apparent in this graph in which up to the scale $r_0$, $D_f\simeq 1.4$. The numerical amount of $r_0$ was obtained by searching a point whose distance from the average slope of its previous points is considerable. For $r>r_0$ the behavior is changed to a new regime in which the graph tends to other linear behavior with some other slope. However since this regime is less than a decade, we can not determine the exact slope of the graph in this region. This graph for general values of $(n_0,R)$ is presented in Fig. \ref{FD2} and Fig. \ref{FD3}. We see that $D_f$ in the first regime is identical for all values of $n_0$ and $R$ and the difference shows itself in the amounts of $r_0$. This shows that for small scales ($r<r_0$) one retrieves nearly the BTW results in which $D_f^{\text{BTW}}=\frac{5}{4}$. One may expect that $r_0$ be an increasing function of $R$. Table [\textbf{I}] and Fig. \ref{FD-scales} show the dependence of $r_0$ on $(n_0,R)$. We see that $r_0$ is increasing function of $R$ and decreasing function of $n_0$ which is obvious in Fig. \ref{FD2} and \ref{FD3}. This shows that for larger values of $n_0$, the second regime dominates the properties of the model, whereas for smaller values, the BTW-type behavior is dominant.\\
The other important quantity is the cluster mass. It is defined as the number of sites involved in a avalanche. Figure \ref{mass} shows the distribution of this function in which it is seen that $N(\text{mass})\sim \text{mass}^{-\tau_m}$. For small values of R $\tau_m=1.1\mp 0.1$ which is consistent with the BTW counterpart. For larger values however, in some point the graph enters a new regime in which $\tau_m=1.4\mp 0.1$. The error bar for determining the cross over point is large due to the smoothness of the cross over. We conclude that for large connection ranges, the system is mainly described by the exponent $\tau_m=1.4$. It is notable that $\tau_m$ is nearly the same for all rates of $n_0$. We end this section with the conclusion that the BTW-type behaviors are dominant for small values of $n_0$ and $R$, whereas for large values of these parameters a new regime is seen in which $\tau_m\simeq 1.4$.  
\section{The NUN as a Time Dependent Process}\label{NUN}
\begin{figure}
\centerline{\includegraphics[scale=.35]{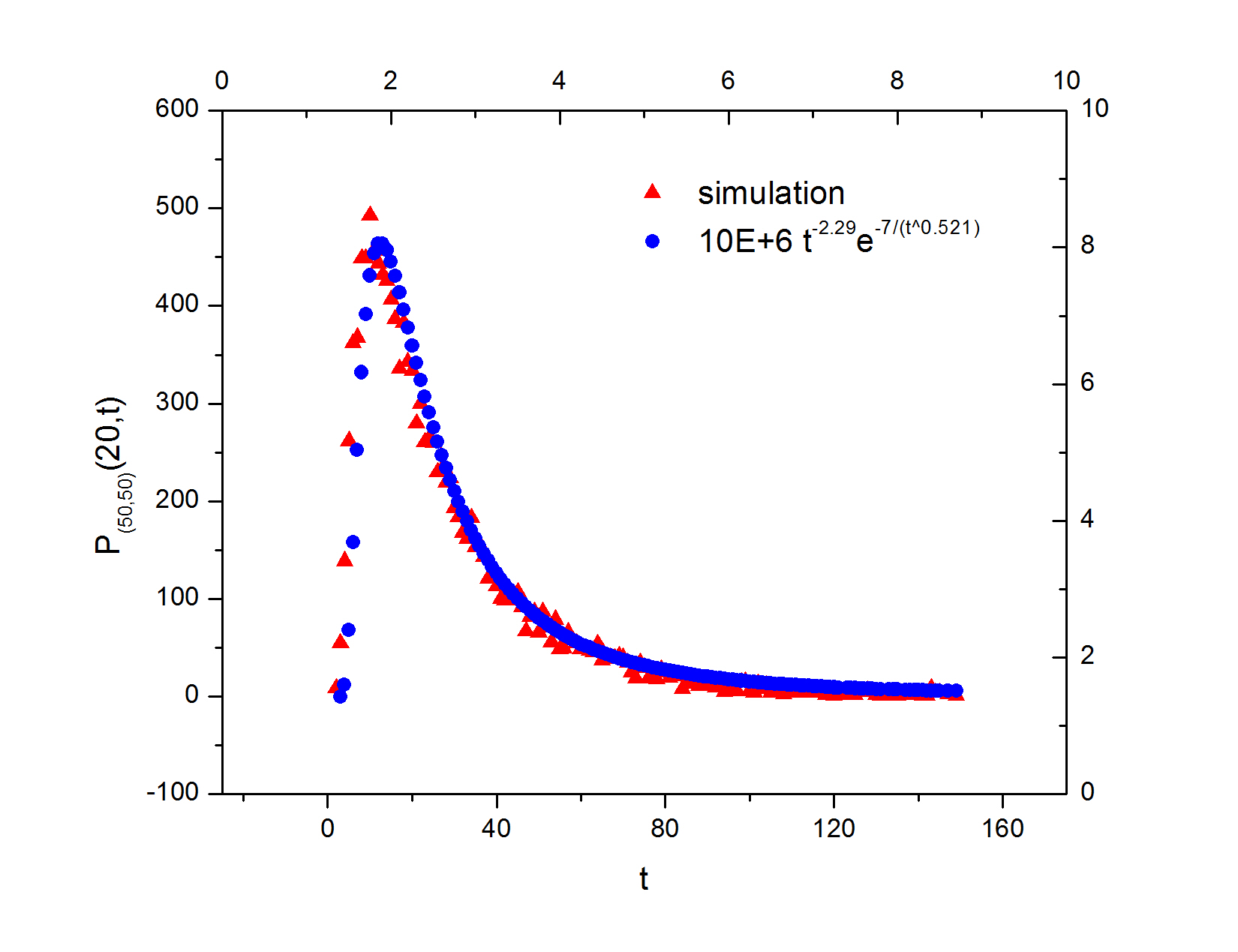}}
\caption{The dependence of $P_{(50,50)}(n=20,t)$ on $t$. (b) The dependence of $\log\left( P_{(50,50)}(n,t=20)\right)$ on $n$.}
\label{t-dep}
\end{figure}

\begin{figure}
\centerline{\includegraphics[scale=.35]{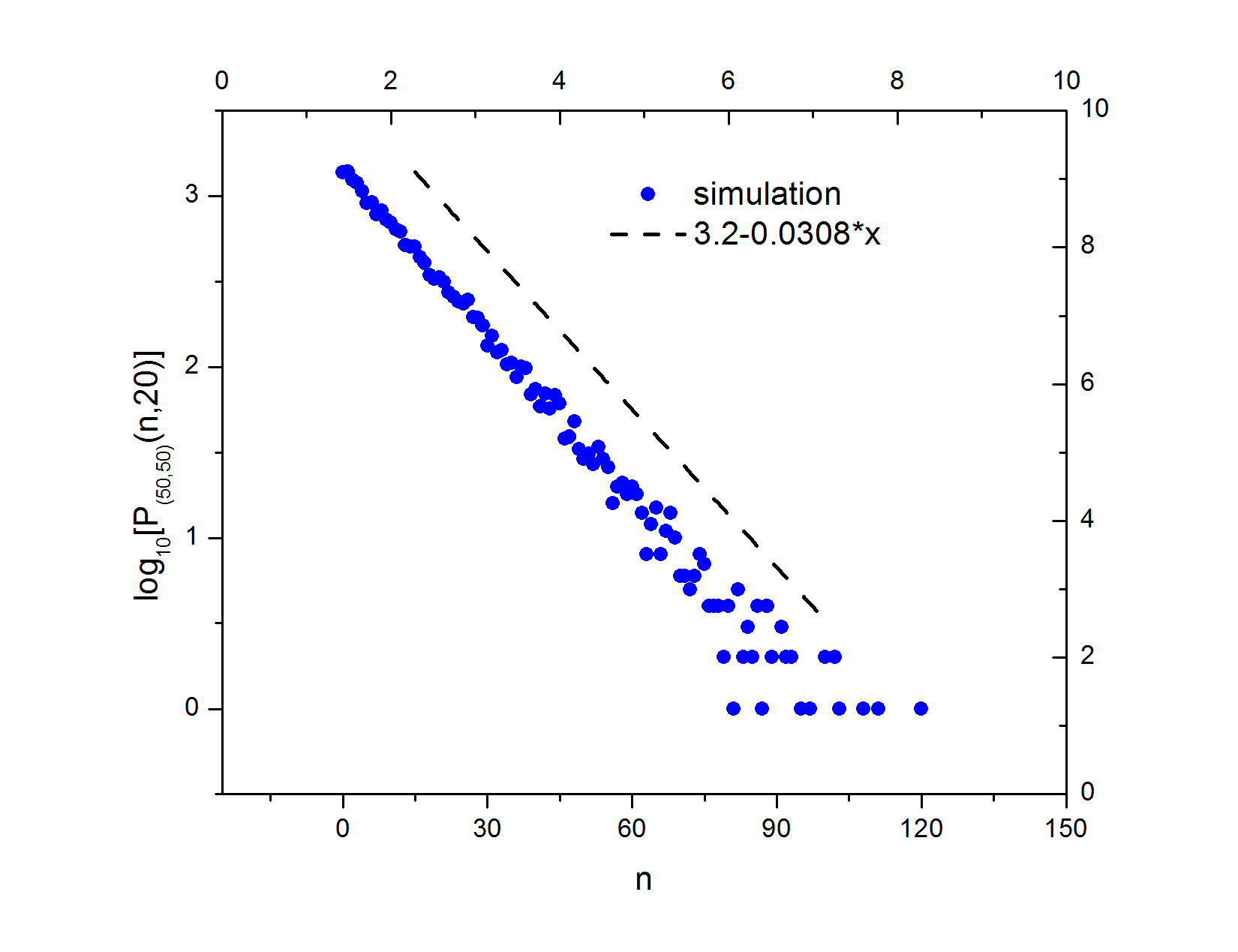}}
\caption{The dependence of $\log\left( P_{(50,50)}(n,t=20)\right)$ on $n$.}
\label{n-dep}
\end{figure}

\begin{figure}
\centerline{\includegraphics[scale=.35]{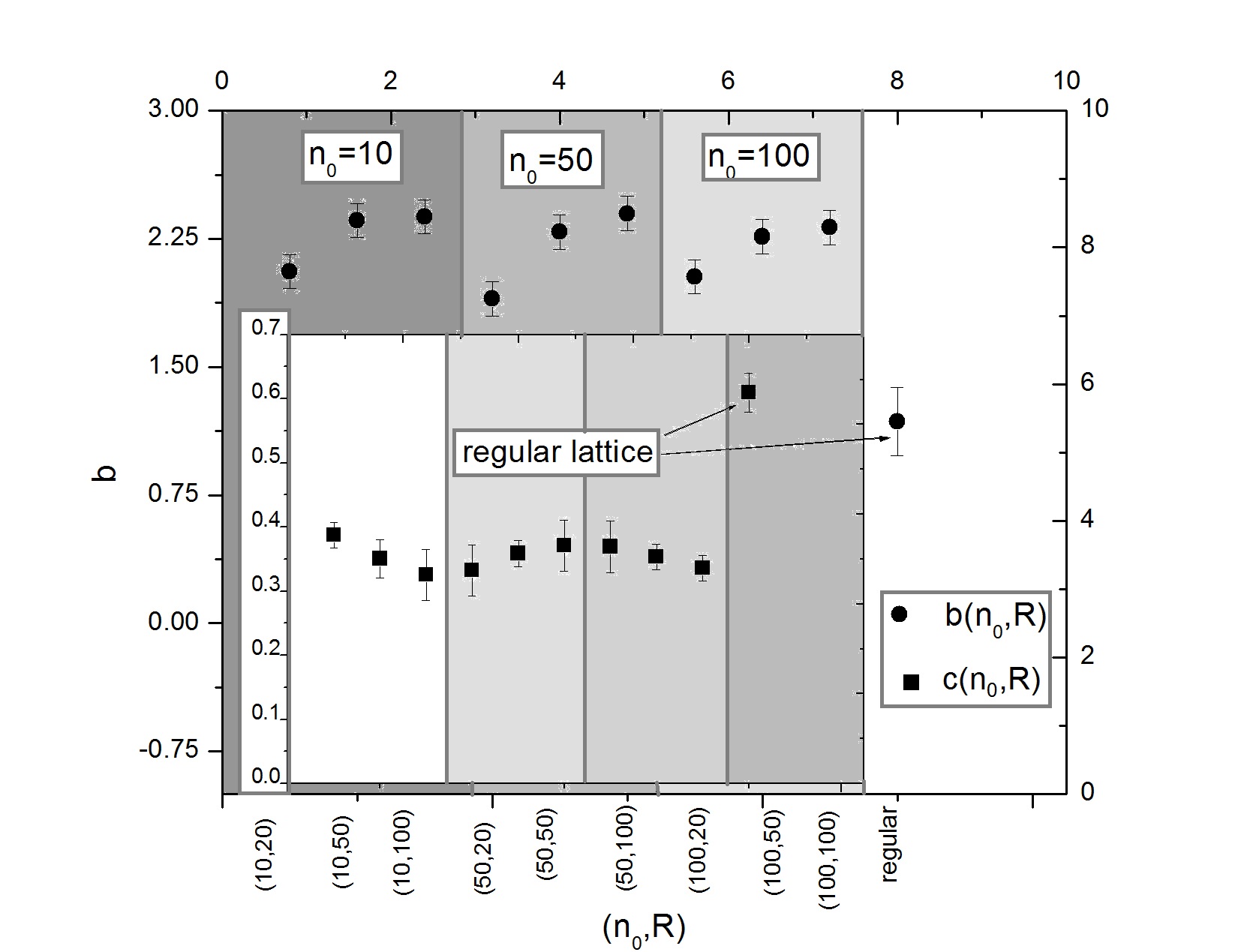}}
\caption{The dependence of the fitting parameters $b$ and $c$ on the parameters $n_0$ and $R$.}
\label{bc-dep}
\end{figure}

\begin{figure}
\centerline{\includegraphics[scale=.35]{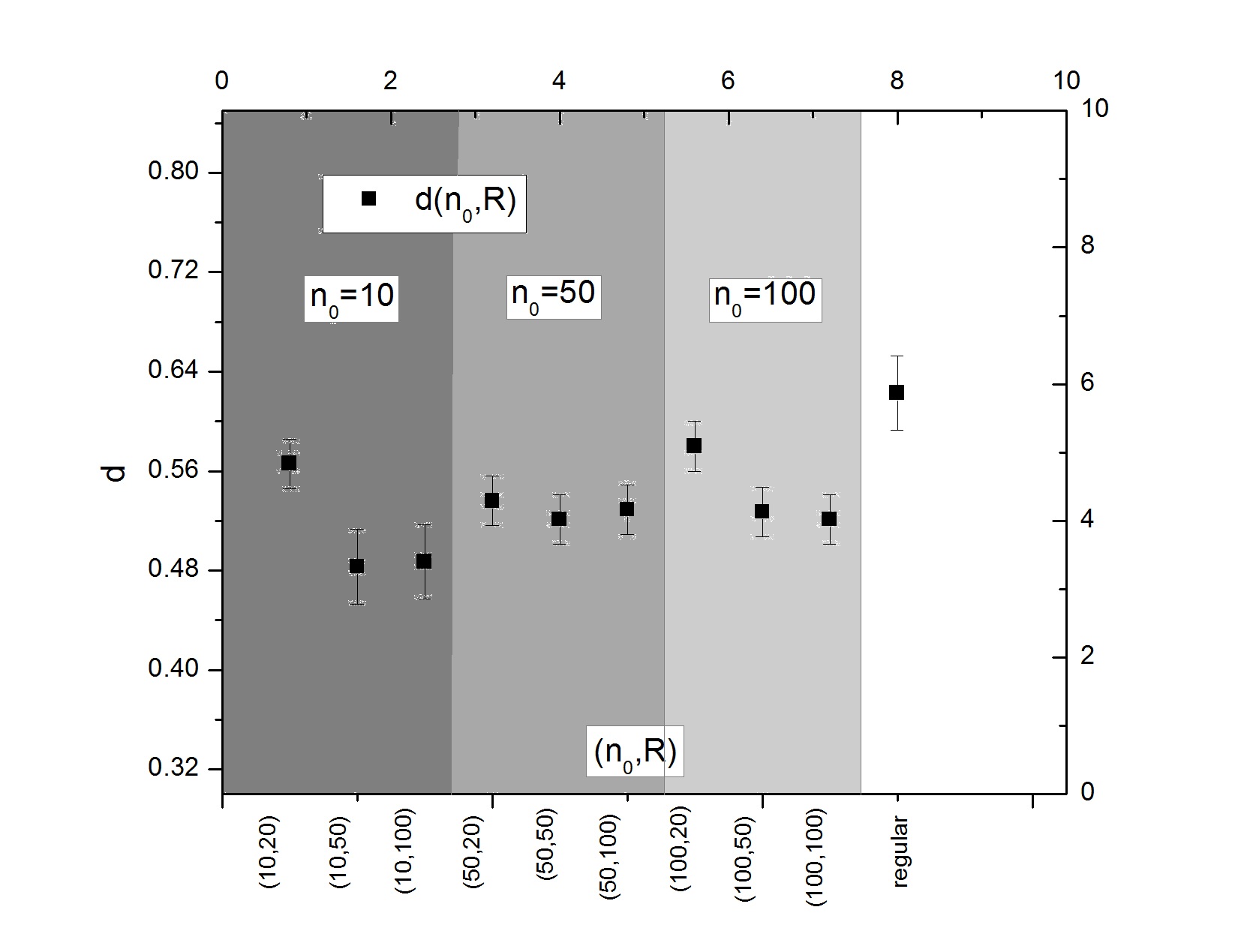}}
\caption{The dependence of the fitting parameter $d$ on the parameters $n_0$ and $R$ (with the error $\mp 0.05$).}
\label{d-dep}
\end{figure}

This section is devoted to the analysis of the NUN as a time dependent stochastic process. In a relaxation process, each $L^2$ stableness check of nodes is defined as a time step (during which each site of the lattice is surely checked once for the toppling) and the NUN is defined as the number of unstable sites to be relaxed in that time. Let us define $P(n,t)$ the probability of having $n$ unstable nodes in time $t$. We found that:
\begin{equation}
P(n,t)=at^{-b}e^{-c\frac{n}{t^d}}
\label{P}
\end{equation}
in which $a,b,c,d$ are the fitting parameters. In Figs. \ref{t-dep} and \ref{n-dep} we have shown $P_{(n_0,R)}$ as an example for the cases $n=\text{const.}=50$ ($t$-dependence) and $t=\text{const.}=50$ ($n$-dependence). The important feature is that $a$, $b$ and $c$ depends on $(n_0,R)$ and $d\simeq 0.5$ is nearly independent of them for random and regular lattices. This dependence is indicated in Figs. \ref{bc-dep} and \ref{d-dep} which shows $c=0.35\mp 0.05$ and $b$ is nearly constant in $n_0$ ($2<b<2.5$ for all ($n_0,R$)) and increasing function of $R$ (except for the regular lattice in which $b\simeq 1$) and $d=0.5\mp0.05$ for both regular and random lattices. Setting $d=0.5$ in Eq. (\ref{P}) we obtain the following symmetry of NUN process:

\begin{equation}
\left\lbrace \begin{array}{c} n\longrightarrow \lambda n \\ t\longrightarrow \lambda^2 t \end{array}\right. \Rightarrow P\longrightarrow \lambda^{-2b}P.
\label{symmetry}
\end{equation}

One can easily check that the Fokker-Planck equation governing $P$ is ($d=0.5$)
\begin{equation}
\partial_TP=-\partial_x\left[ (1-b)P\right]+\frac{1}{2}\partial_x^2\left[ xP\right]
\end{equation}
in which $T\equiv \frac{2}{c}\sqrt{t}$ and $x$ is the continuum limit of $n$. This equation respects the symmetry Eq. (\ref{symmetry}) and leads to the following Langevin equation:
 \begin{equation}
dx=(1-b)dT +\sqrt{x}dW(T)
\label{n-process}
\end{equation}
\begin{figure}
\centerline{\includegraphics[scale=.40]{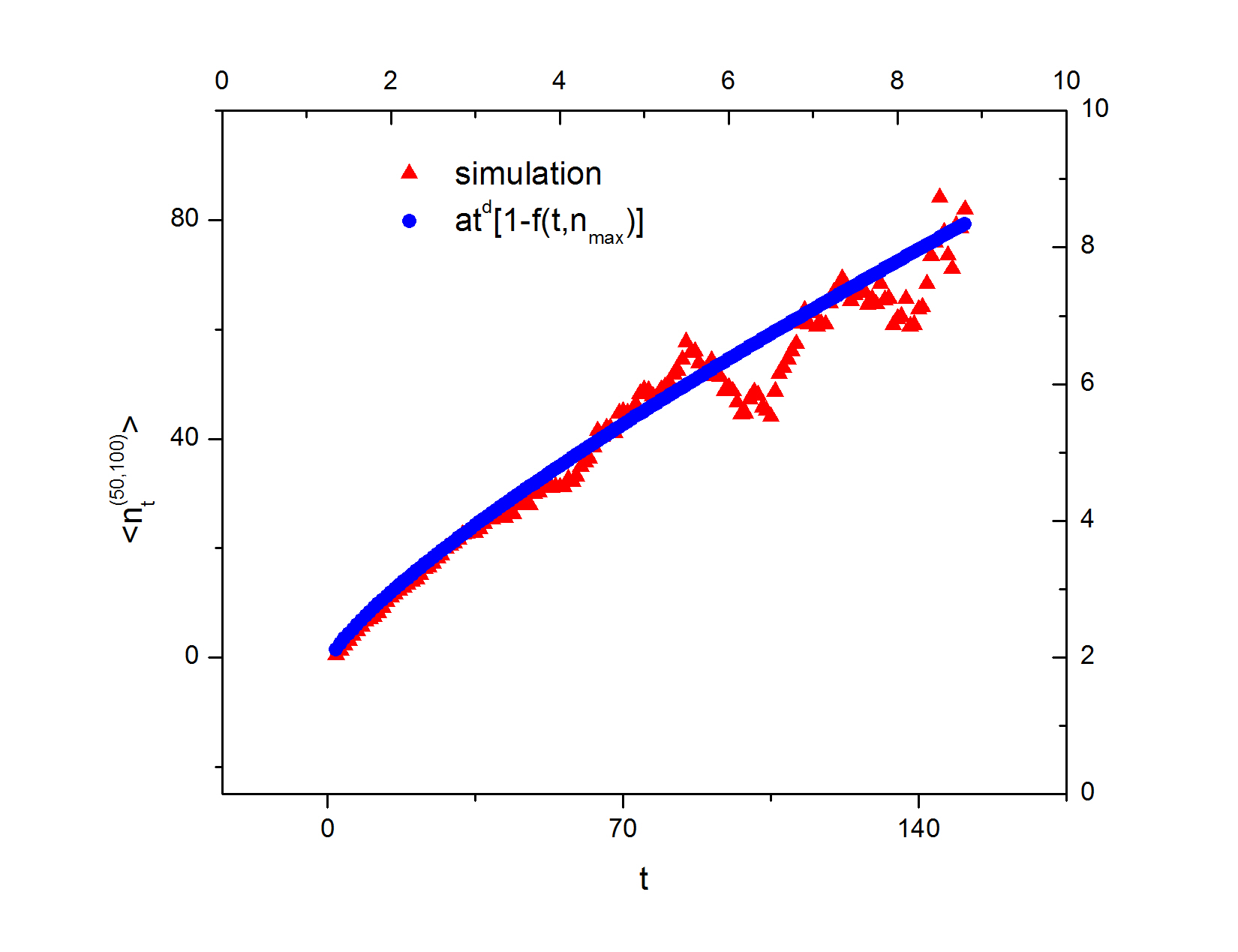}}
\caption{The plot of $\left\langle n(t)\right\rangle $ versus $t$ in which $f(t,n_{\text{max}})=\frac{cn_{\text{max}}}{t^d}\frac{e^{-\frac{cn_{\text{max}}}{t^d}}}{1-e^{-\frac{cn_{\text{max}}}{t^d}}}$.}
\label{n_aver}
\end{figure}
in which $W(T)$ is a one dimensional Wiener process. For regular lattice $b\simeq 1$ from which one realizes that $x$ (up to a normalization) is proportional to a Wiener process, i.e., $dx=\sqrt{x}dW(T)$. As soon as the randomness in lattice links is "turned on", $b$ jumps from $1$ to another values and $x$ acquires a drift term according to Eq. (\ref{n-process}). The Eq. (\ref{n-process}) for $b=1$ can be reformed, by transformation $y\equiv 2\sqrt{x}$, to:
\begin{equation}
dy=dW(T)-\frac{1}{2y}dT
\end{equation}
which is a zero-dimensional Bessel process. We can also calculate $\left\langle n_t\right\rangle$ by integrating $P(n,t)$:
\begin{equation}
\left\langle n_t\right\rangle = \frac{\int _0^{n_{\text{max}}}nP(n,t)dn}{\int _0^{n_{\text{max}}}P(n,t)dn}=\frac{t^d}{c}\left( 1-\frac{cn_{\text{max}}}{t^d}\frac{e^{-\frac{cn_{\text{max}}}{t^d}}}{1-e^{-\frac{cn_{\text{max}}}{t^d}}}\right).
\label{n_ave}
\end{equation}
in which the second term is deduced from Eq. (\ref{P}) and $n_{\text{max}}$ is the upper bound of $n$ in our analysis. Figure \ref{n_aver} demonstrates the agreement between our numerical results and Eq. (\ref{n_ave}) for $d=0.6$ and $cn_{max}=1914$. The important feature of this relation is that for small times $n_t\sim t^d$ in which $d$ lies within the interval demonstrated in Fig \ref{d-dep}. This may be expected from the symmetry Eq. (\ref{symmetry}).
\begin{acknowledgements}
I wish to thank P. Manshour for his useful hints on this work.
\end{acknowledgements}
\section{Conclusion}
 We have considered the BTW model in random link lattices with finite range interaction (RLFRI) for which the degree distribution for nodes is considered to be uniform in the interval $(0,n_0)$ and interaction range is a finite value ($R$). As well as the mass and loop length, we analyzed the Dijkstra gyration radius statistics for the model in hand. We found a length scale in which the BTW-type behaviors of the model is changed to a new one. NUN as a time process has been also investigated in this analysis. It has been shown that it has identical symmetry for regular and random lattices, but its exact form is different for two cases. We obtained that it is (up to a normalization) proportional to a one dimensional Weiner process for regular lattice and for RLFRI it acquires a drift term.

\end{document}